# Verification Code Recognition Based on Active and Deep Learning

Dongliang Xu[1*], Bailing Wang[2], XiaoJiang Du[3], Xiaoyan Zhu[4], zhitao Guan[5], Xiaoyan Yu[6], Jingyu Liu[7]

1 School of Computer Science and Technology, Shandong University, Weihai, P.R. China
2 School of Computer Science and Technology, Harbin Institute of Technology, Weihai, P.R. China
3 Department of Computer and Information Sciences, Temple University, Philadelphia, PA, USA
4 Computer information center, Xidian University, Xi'an, P.R. China
5 School of Control and Computer Engineering, North China Electric Power University, Beijing, P.R. China
6 Dept. of Computer Science and Technology, Capital Normal University, Beijing, P.R. China
7 Computer information center, Intermediate People's Court, Weihai, P.R. China
xudongliang@sdu.edu.cn

*Abstract*—A verification code is an automated test method used to distinguish between humans and computers. Humans can easily identify verification codes, whereas machines cannot. With the development of convolutional neural networks, automatically recognizing a verification code is now possible for machines. However, the advantages of convolutional neural networks depend on the data used by the training classifier, particularly the size of the training set. Therefore, identifying a verification code using a convolutional neural network is difficult when training data are insufficient. This study proposes an active and deep learning strategy to obtain new training data on a special verification code set without manual intervention. A feature learning model for a scene with less training data is presented in this work, and the verification code is identified by the designed convolutional neural network. Experiments show that the method can considerably improve the recognition accuracy of a neural network when the amount of initial training data is small.

## I. INTRODUCTION

A verification code is a publicly automated program used to distinguish whether a user is a computer or a human [1-3]. Such program must generate tests that humans can easily pass but are difficult for computers to accomplish. The aim of verification code identification algorithm is to automatically identify the content in verification code which can better improve work efficiency and ensure real-time [4-6]. The verification code generation technology can be promoted by the study of verification code identification technology. Each of the verification code generation or recognition progress will affect the safety of verification code, this will also promote more advanced verification code technology and rapid development of the image recognition technology [7-9].

Verification code identification technology is widely researched abroad especially in developed countries. A valid verification code should have a human recognition rate higher than 80%, and the probability that a machine will recognize its content by using certain resources should be less than 0.01% [10,11]. Verification code technology has been widely used to improve the security and anti-attack capabilities of websites due to its simplicity, easy implementation, and small amount of data transmission. This technology aims to prevent the bulk registration of websites, violent password breaking, and malicious attacks.

Researchers have begun to study techniques for automatically identifying verification codes. Most concepts for these solutions include performing character segmentation and identification. However, this type of scheme has limitations for verification codes that contain skewed characters. Therefore, a rectangular window cannot be used to split characters, and a more efficient classification method is necessary. In addition, a verification code picture differs from an object in a natural image, as shown in Figure 1. Verification code pictures and training data are relatively different due to certain factors, such as skewness of characters.

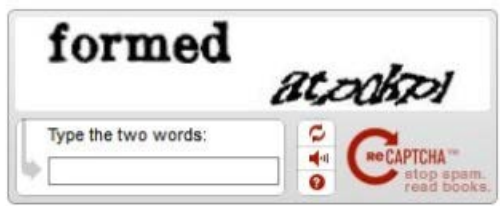

Figure 1 Sample of a verification code

This study proposes a method based on active learning to solve the aforementioned problems. First, a small-scale set is used for training. In each training process, the addition of samples to the training data is determined according to the uncertainty associated with the classifier and the given prediction result. Assuming that the uncertainty is calibrated, the information-rich samples are selected for training, and the number of training samples required for the algorithm is considerably reduced compared with that for standard passive learning. In addition, the algorithm is applicable to situations wherein the input data considerably differ from the current training data. Unlike other forms of active learning, the algorithm does not require human supervision to obtain real-time training tags. Instead, the return value is automatically obtained when verification code recognition is used. The current data are used for retraining when the classifier correctly identifies the verification code. At this moment, the real-time training tag is known. When recognition fails, no similar sample exists in the training set, and the algorithm will demonstrate the training of a neural network only through successfully identified samples. In summary, this research has the following objectives:

1. To calculate the uncertainty associated with the given predictor results and to determine the relationship to the correct classification.

2. To use a convolutional neural network for active learning.

3. To conduct training with correct but uncertain classification samples and obtain good training results with less training data in the absence of manual intervention.

II. RELATED WORK

The traditional method for detecting text in an image includes two steps [12]: positioning and identifying the words or characters in an image. Most domestic researches have focused on simpler codes, such as digital and English ones which are more normal and with no distortion deformation, but some domestic researches have started to research characters with distorted adhesion and whose in background and serious noise.Liu et al. [13] propose an anisotropic heat kernel equation group which can generate a heat source scale space during the kernel evolution based on infinite heat source axiom, design a multi-step anisotropic verification code identification algorithm which includes core procedure of building anisotropic heat kernel, setting wave energy information parameters, combing out verification code characters and corresponding peripheral procedure of gray scaling, binarizing, denoising, normalizing, segmenting and identifying, give out the detail criterion and parameter set. Actual test show the anisotropic heat kernel identification algorithm can be used on many kinds of verification code including text characters, mathematical, chinese, voice, 3D, programming, video, advertising, it has a higher rate of 25% and 50% than neural network and context matching algorithm separately for Yahoo site, 49% and 60% for Captcha site, 20% and 52% for Baidu site, 60% and 65% for 3DTakers site, 40% and 51% for MDP site.

Wen Mingli [14] et al. proposed a depth neural network model to identify the verification code, the introduction of convolution neural network, the verification code without any pretreatment, end-to-end identification verification code, to avoid the traditional positioning, segmentation and other steps, simulation training verification code data, expand the network required data sets, improve the accuracy of verification code recognition. Bao Qian [15] et al. used Hopfield neural network (Hopfield neural network model is as Figure 2) to identify verification code, which solved the problem of transparent integration in system development, and provide users with the perfect interface. And LeCun et al. [16] proposed using convolutional neural networks to identify handwritten numbers, which are hierarchically constructed to classify objects in a certain order.

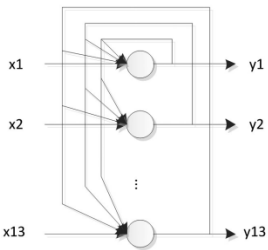

Figure 2 Hopfield neural network model

Goodfellow et al. [17] proposed the use of deep convolutional neural networks, which are widely applied in other fields, such as image classification, speech recognition, and natural language processing. Training a convolutional neural network requires a massive training set. However, the available labeled training dataset for verification code identification is small. In addition, a dictionary can be used to exclude impossible words. For example, Mori and Malik [18] proposed a method for using a dictionary to identify a verification code. The method are based on shape context matching that can identify the word in an EZ-Gimpy image with a success rate of 92% and the requisite 3 words in a Gimpy image 33% of the time. Chellapilla and Simard [19] initially separated a verification code into individual characters and then reidentified the code without using a dictionary. However, given that the characters in a verification code currently overlap each other, simply splitting a single character using a rectangular window is impossible. These verification codes are similar to handwritten text. Jaderberg et al. [20] developed a convolutional neural network for identifying text in natural scene images. However, the authors must manually create a massive set of text images to achieve training results. By contrast, the proposed algorithm uses a smaller training set and applies active learning to fine-tune a neural network at runtime, thereby obtaining correct classification results and test samples with high uncertainty. Several papers [21-23] have studied the related issues.

III. OVERVIEW OF THE PROPOSED APPROACH

A. *Deep convolutional neural network*

This study proposes a deep convolutional neural network for identifying an entire sequence of verification codes without presegmenting the images. The neural network structure used is shown in Figure 3.

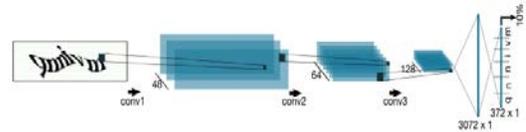

Figure 3 Structure of a convolutional neural network

Assume that in a six-digit verification code, each bit is represented by 62 neurons of the output layer. A bijective function $\theta(x)$ is defined, and a character $x \in \{'0', \ldots '9', 'A', \ldots 'Z', 'a', \ldots 'z'\}$ is mapped onto an integer $l \in \{0 \ldots 61\}$:

$$\theta(x) = \begin{cases} 0 \ldots 9, & x = {'0'} \ldots {'9'} \\ 10 \ldots 35, & x = {'A'} \ldots {'Z'} \\ 36 \ldots 61, & x = {'a'} \ldots {'z'} \end{cases} \quad (1)$$

The first 62 neurons are assigned to the first character of the sequence, and then the last 62 neurons are assigned to the second character, and so on, in a similar manner. For any character $x_i$, the corresponding neuron n is calculated as $n = i \times 62 + \theta(x_i)$, $i \in \{0 \ldots 5\}$, i represents the number of bits, and the output layer contains $6 \times 62 = 372$ neurons. When predicting a character, the corresponding 62 neurons should be considered, assuming that their sum is 1. Figure 3 and 4 shows the output of a neural network, which is the index of the predicted first character $c_0 = 52$, and its corresponding prediction label is $x = \theta^{-1}(c_0) = 'q'$.

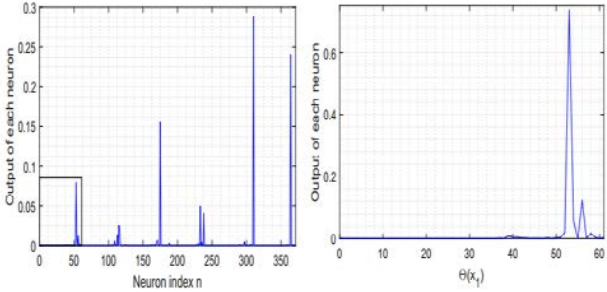

Figure 4 Sample of a neural network output

### B. Active learning for reducing the amount of training data

Convolutional neural networks typically require an extremely large training set to facilitate accurate classification. However, collecting verification codes for millions of human markers is infeasible; hence, this study proposes the use of active learning (Figure 5). The main idea is to add new training data only when necessary. If the amount of information in the sample is sufficient for retraining, then no new datum is added. This condition is determined by the uncertainty of the prediction results, which are calculated based on the optimal label and suboptimal labeling strategy described later in this paper.

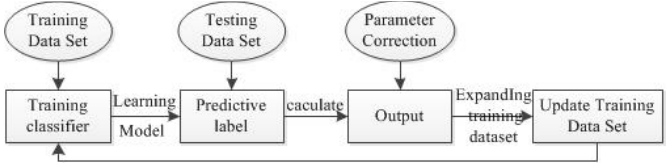

Figure 5 Active learning processing

1) Calculation of uncertainty

As described earlier, the distribution of predicted values for each character is estimated by setting the sum of the outputs of the respective neural networks as 1. The total uncertainty and the suboptimal labeling strategy are then used to calculate the total uncertainty.

$$\eta = \frac{1}{d} * \sum_{i=1}^{d} \frac{\arg\max\{P(x_i)/\arg\max P(x_i)\}}{\arg\max P(x_i)} \quad (2)$$

Where $P(x_i)$ is the sum of the neural network outputs for the character $d_i$, and the suboptimal value is divided by the best prediction for each character.

2) Obtaining measured information

The algorithm uses only data samples with correct tags for retraining, learns without human intervention, and successfully identifies the verification code to obtain the correct text content. However, retraining with only these correctly classified samples is highly inefficient. Given that the classifier will perform better over time, the correct rate of classifying the sample will increase and training will become more frequent. Therefore, a neural network should be retrained each time a new sample is correctly classified. To avoid this process, the algorithm uses the uncertainty values calculated earlier, in which the correctly classified samples are sorted by the uncertainty value of the predicted results in each training cycle, and only the most uncertain test samples are used for retraining. This approach reduces the required number of training samples, but the most uncertain samples in the training process are the most informative, as demonstrated in the experiments.

## IV. EXPERIMENTAL AND ANALYSIS

This research conducts experiments on an automatically generated set of verification codes using the TensorFlow deep learning framework.

### A. Dataset Generation

Given that no manually marked verification code dataset is available, this experiment uses a script to generate a verification code. During automatic generation, the absence of duplicate verification codes in the dataset must be ensured. The verification code set is generated using the PHP verification code frame. The verification code consists of a slanted text with a fixed length of 6, similar to Google's reCAPTCHA, with a size of 180 × 50. The framework is modified to generate black and white images. In addition, shadows and lines are disabled in the text, whereas random characters are used. No dictionary vocabulary is used, and thus, the rule that the second character must be a vowel is disregarded, and the font is modified to "AntykwaBold." The automatically generated verification code is shown in Figure 6.

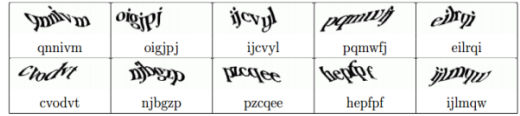

Figure 6 Verification code samples used in the experiment

### B. Neural Network Design

This experiment uses a convolutional neural network (Figure 3). The convolutional layers are 48, 64, and 128 in size, with a kernel size of 5 × 5 and a fill size of 2. The pooling layer window size is 2 × 2. The first and third pooling layers and the first convolutional layer have a step size of 2. The neural network has a fully connected layer with a size of 3072, and the output of the second fully connected layer (classifier) is 372. After each convolutional layer and the first fully connected layer, a linear correction unit and a dropout are added. The batch size for each iteration is 64.

### C. Qualitative Analysis

The neural network is trained using the stochastic gradient descent algorithm. The neural network trained in this experiment independently identifies all characters of the verification code compared with other methods. The rule for learning rate change is $\alpha = \alpha_0 \times (1 + \gamma \times t)^{-\beta}$, the initial learning rate $\alpha_0 = 10^{-2}$, $\beta = 0.75$, $\gamma = 10^{-4}$, and t is the number of current iterations. Momentum μSet as $\mu = 0.9$, and regularization parameter λ is $5 \times 10^{-4}$.

Obtaining a large training sample set is difficult; thus, the algorithm aims to reduce the size of the initial training set. First, the network is trained using an extremely small initial training set. The initial training set is $10^4$ images, and $5 \times 10^4$ iterations are performed. The accuracy is only 9.6% and even decreases with an increase in the number of iterations. Therefore, the algorithm must use active learning.

First, the neural network is trained using $10^4$ training images for $5 \times 10^4$ iterations. Then, $5 \times 10^4$ test images are classified. A new training sample is selected from the correctly classified samples. At this point, either all the samples or only $5 \times 10^3$ samples are selected based on their uncertainty to make the highest, lowest, or random choice depending on uncertainty. The calculation of uncertainty is described in Section 4.1. Once the newly selected sample is added to the training set, the network will be retrained for $5 \times 10^4$ iterations. Then, the same process is performed. The algorithm is applied to the 20 round active learning process. Accuracy is calculated once for each $5 \times 10^3$ iteration on a fixed training set. Finally, a neural network with correct prediction results but with uncertainty is obtained. All the results adopt the average of two runs.

Increasing the number of samples in the training set requires additional storage space. Moreover, increasing the number of iterations can improve the training effect but will increase training time. Therefore, the network is retrained using only selected samples in each iteration. $10^4$ training images for $5 \times 10^4$ iterations are initially used. Then, $10^5$ test images are classified. The training set that contains $10^4$ initial images is replaced with the correctly classified image set, and $2.5 \times 10^4$ iterative training is conducted. The number of iterations after each round of training is reduced according to the following rules: $2.5 \times 10^4$ times, round 6; $2 \times 10^4$ times, round 11; $1.5 \times 10^4$ times, round 16; $1 \times 10^4$ times, round 21; $5 \times 10^3$ times, until round 40. Finally, the neural network with the predicted result is correct but still exhibits uncertainty. Therefore, performance is optimized. Correctly classifying a neural network is reasonable, but the result exhibits uncertainty. Therefore, learning can be achieved from a correctly classified collection. Learning with a misclassified sample produces better results but is infeasible in practice.

V. CONCLUSION

This study proposes a verification code recognition technique that initially uses a small set of images to train deep convolutional neural networks and then adopts test samples to improve the classifier. A new sample is selected from the test set to be included in the training set based on the uncertainty of the test sample. The experimental results show that the use of well-classified but uncertain test samples can considerably improve the performance of a neural network.

ACKNOWLEDGMENT

This research was partially supported by the National Natural Science of China under grants 61371177 and U1431102, Shandong Provincial Natural Science Foundation under grant ZR2015AM015, Science and Technology Major Project of Shandong under grant 2015ZDXX0201B04, Science and Technology Development Program of Shandong Province under grant 2014GGX101053 and 2018GGX101023, and Science and Technology Development Program of Weihai Province under grant 2014GGX101053.

REFERENCES

[1] LeCun Y, Boser B, Denker J S, et al. Backpropagation applied to handwritten zip code recognition[J]. Neural computation, 1989, 1(4): 541-551.
[2] Hou X, Niu Z, Ben G. Method and device for realizing verification code: U.S. Patent Application 15/636,453[P]. 2017-10-19.
[3] Q. Yang, J. Yang, W. Yu, D. An, N. Zhang and W. Zhao, "On False Data-Injection Attacks against Power System State Estimation: Modeling and Countermeasures," in IEEE Transactions on Parallel and Distributed Systems, vol. 25, no. 3, pp. 717-729, March 2014.
[4] El Ahmad A S, Yan J, Ng W Y. CAPTCHA design: Color, usability, and security[J]. IEEE Internet Computing, 2012, 16(2): 44-51.
[5] Jiang N, Dogan H. A gesture-based captcha design supporting mobile devices[C]//Proceedings of the 2015 British HCI Conference. ACM, 2015: 202-207.
[6] X. Du, Y. Xiao, M. Guizani, and H. H. Chen, "An Effective Key Management Scheme for Heterogeneous Sensor Networks," Ad Hoc Networks, Elsevier, Vol. 5, Issue 1, pp 24–34, Jan. 2007.
[7] Yang L, Gurumani S, Fahmy S A, et al. Automated Verification Code Generation in HLS Using Software Execution Traces[C]//Proceedings of the 2016 ACM/SIGDA International Symposium on Field-Programmable Gate Arrays. ACM, 2016: 278-278.
[8] Huang C, Xiong L, Peng W, et al. NUMERICAL VERIFICATION CODE GENERATION METHOD AND DEVICE: U.S. Patent Application 15/542,196[P]. 2018-6-14.
[9] X. Du, M. Guizani, Y. Xiao and H. H. Chen, Transactions papers, "A Routing-Driven Elliptic Curve Cryptography based Key Management Scheme for Heterogeneous Sensor Networks," IEEE Transactions on Wireless Communications, Vol. 8, No. 3, pp. 1223 - 1229, March 2009.
[10] Bengio, Y., Ducharme, R., Vincent, P., Janvin, C.. A neural probabilistic language model. The Journal of Machine Learning Research 3, 1137{1155 (2003)
[11] Y. Cheng, X. Fu, X. Du, B. Luo, M. Guizani, "A lightweight live memory forensic approach based on hardware virtualization," Vol. 379, pp. 23-41, Elsevier Information Sciences, Feb. 2017.
[12] Bai J, Chen Z, Feng B, et al. Chinese image text recognition on grayscale pixels[C]//Acoustics, Speech and Signal Processing (ICASSP), 2014 IEEE International Conference on. IEEE, 2014: 1380-1384.
[13] Lizhao L, Jian L I U, Yaomei D, et al. Design and implementation of verification code identification based on anisotropic heat kernel[J]. China Communications, 2016, 13(1): 100-112.
[14] Wen M, Zhao X, Cai M. An end-to-end verification code identification based on depth learning[J]. Wireless Internet Technology, 2017.
[15] Bao Q, Li W, Zhang Q. Research on Recognition of the Verification Code Based on Android[J]. Bulletin of Science & Technology, 2017.
[16] Lecun, Y., Bottou, L., Bengio, Y., Haffner, P. Gradient-based learning applied todocument recognition. Proceedings of the IEEE (1998)
[17] Goodfellow, I.J., Bulatov, Y., Ibarz, J., Arnoud, S., Shet, V. Multi-digit number recognition from street view imagery using deep convolutional neural networks.ICLR (2014)
[18] Mori G, Malik J. Recognizing objects in adversarial clutter: Breaking a visual CAPTCHA[C]//Computer Vision and Pattern Recognition, 2003. Proceedings. 2003 IEEE Computer Society Conference on. IEEE, 2003, 1: I-I.
[19] Von Ahn L, Blum M, Hopper N J, et al. CAPTCHA: Using hard AI problems for security[C]//International Conference on the Theory and Applications of Cryptographic Techniques. Springer, Berlin, Heidelberg, 2003: 294-311.
[20] Jaderberg, M., Vedaldi, A., Zisserman, A. Deep features for text spotting. In: CVPR (2014)
[21] L. Wu, X. Du, and J. Wu, "Effective Defense Schemes for Phishing Attacks on Mobile Computing Platforms," IEEE Transactions on Vehicular Technology, Issue 8, Vol. 65, pp. 6678 - 6691, June 2016.
[22] W. Yu et al., "A Survey on the Edge Computing for the Internet of Things," in IEEE Access, vol. 6, pp. 6900-6919, 2018.
[23] X. Hei, X. Du, S. Lin, and I. Lee, "PIPAC: Patient Infusion Pattern based Access Control Scheme for Wireless Insulin Pump System," in Proc. of IEEE INFOCOM 2013, Turin, Italy, Apr. 2013.